# Scanning diffraction imaging without stable illumination and scan position information

Tao Liu,[1] Bingyang Wang,[1] Jiangtao Zhao,[2] Maik Kahnt[3], Fucai Zhang[1,*]

[1]Department of Electrical and Electronic Engineering, Southern University of Science and Technology, 1088 Xueyuan Avenue, Shenzhen 518055, China
[2]ESRF – The European Synchrotron, Grenoble, France
[3]MAX IV Laboratory, Lund University, 22100 Lund, Sweden
* Corresponding authors: zhangfc@sustech.edu.cn

**Abstract**
Ptychography has gained significant prominence at synchrotron facilities globally for characterizing the topological structure and intricate properties of biological and material specimens at the nanometer or atomic scale, owing to its lens-less, highly quantitative phase imaging capabilities. Its high-resolution imaging potential heavily relies on accurate lateral scan position information, substantial overlap ratio, and highly stable probe function. However, as investigations progress to atomic scales, these requirements become increasingly challenging to meet, often necessitating the integration of high-precision motion control systems and sophisticated manual pre-calibration. In this paper, we propose an innovative imaging framework that eliminates the necessity of those strict requirements of traditional ptychography through the insertion of a wavefront modulator and a new phase retrieval workflow. Notably, our method does not necessitate pre-calibration of the wavefront modulator. Optical validation demonstrated that the deviation between the recovered position and a state-of-the-art motion stage was less than 10 nm. Furthermore, sub-pixel position accuracy was still achievable even with datasets exhibiting only a 13% overlap ratio. For our X-ray experiment with a spatially unstable probe violating the ptychographic imaging model and 18% overlap ratio, our method successfully retrieved beam spatial deviation, scan position, probe function, and sample transmission function simultaneously and quantitatively. Importantly, the probe spatial deviation of 500nm along the X-axis and 50nm along the Y-axis were measured, which were not accessible by previous ptychographic methods. With these experimentally verified promising features, we expect the proposed method to be a powerful imaging method that features auto-quantitative calibration of various experimental parameters; it will greatly simplify the implementation and widen the scope of applications of scanning diffraction microscopy, especially at nanometer and atomic scale imaging.

## 1. Introduction

Complex-valued transmission functions offer crucial insights into the structure and properties of matter, playing a vital role in biological studies and the understanding of nanocrystals [1-4]. For a wavefield of light whose frequency exceeds the response bandwidth of the detector, the phase information is lost, leaving only the measurable diffraction intensity. Phase retrieval techniques have evolved for decades, attempting to recover the lost phase from the measured intensity [5-8].

Coherent diffraction imaging (CDI), a single-shot phase retrieval technique, relies on oversampled diffraction patterns, prior knowledge of a well-defined sample area (termed support), and an iterative algorithm to recover the sample's exit wave [9-14]. Over the past years, CDI has significantly advanced in studying isolated cells, particles, and nanocrystals [15-19]. However, the requirement of isolated samples limits its applicability to extended samples [20-24].

Ptychography, conversely, enables the recovery of extended objects and shared probe functions using sufficiently overlapped scans, oversampled diffraction patterns, and specific phase retrieval engines [25-30]. The spatial resolution of Ptychography depends on the accuracy of relative position measured by the physical positioning system [31-34]. For high-resolution microscopy systems at the nano or atomic scale, even slight positional uncertainties within a fraction of a pixel can lead to significant degradation of image quality [32-39]. Consequently, considerable research papers regarding the instrumentation and algorithms have been made to guarantee successful reconstruction. Specifically, integrating the interferometer provides accurate prior scan position information [25,39], while algorithmic enhancements such as conjugate gradient descent, genetic algorithm, and correlation-based algorithms further refine the measured position during the reconstruction process [32,34,37-39]. The above techniques have been proved efficient when the error of scan position stays at a tolerable level, and ptychographic engines can still yield a coarse estimate of probe and object functions. However, for Electron Ptychography probing atomic-scale structures, even a coarse estimate of sample and probe function still demands significant efforts due to the demand for constructing a highly precise positioning system and the mechanical vibration [33,38-46].

In traditional ptychographic scheme, a stable probe is assumed to be stable during data acquisition. However, in practice, fluctuations in intensity, phase, and lateral position are inevitable. Although considerable techniques have been proposed to compensate for the fluctuations and improve image quality by statistical methods, dramatic spatial and temporal probe variation can still be challenging for those techniques [42-46,49-51]. Moreover, these methods provide only a statistical assessment of beam quality rather than directly measuring specific physical quantities' deviations.

The above challenges impede the development of microscopy towards a smaller scale, so we propose a novel imaging scheme incorporating features of Ptychography and Coherent Modulation Imaging, which requires no prior knowledge of scan position and less dependence on the stable probe assumption and overlap ratio [25,47]. The apparatus and workflow of the presented data processing procedure are shown in Fig.1. Physically, we introduce a diffraction optical component downstream of the sample to induce strong wavefront modulation in the optical path, which alters the exit waves and generates a diffraction pattern with a specific distribution. Algorithmically, the modulated diffraction intensity distribution enhances the convergence of iterative reconstruction by constraining the convolution spectrum of object function and the modulation function, which enables the object waves of each scan point to be reconstructed independently; with the sample image in real space, the relative position of scan can be calculated by registering the shift of shared features due to the overlapping nature of the scan. For two-dimensional imaging, the information obtained above is sufficient to assemble the completed sample information also conventional ptychographic engine such as the extended ptychographic iterative engine (ePIE) or the difference map (DM) to be implemented [30,48]. When a stable probe is not valid, the independent exit wave recovery makes it possible to measure and compensate the deviation for the deviation of the probe numerically or incorporate probe variation in the imaging model. Optical and X-ray experiments using different types of samples and experiment parameters were carried out to validate the different capabilities of the presented method.

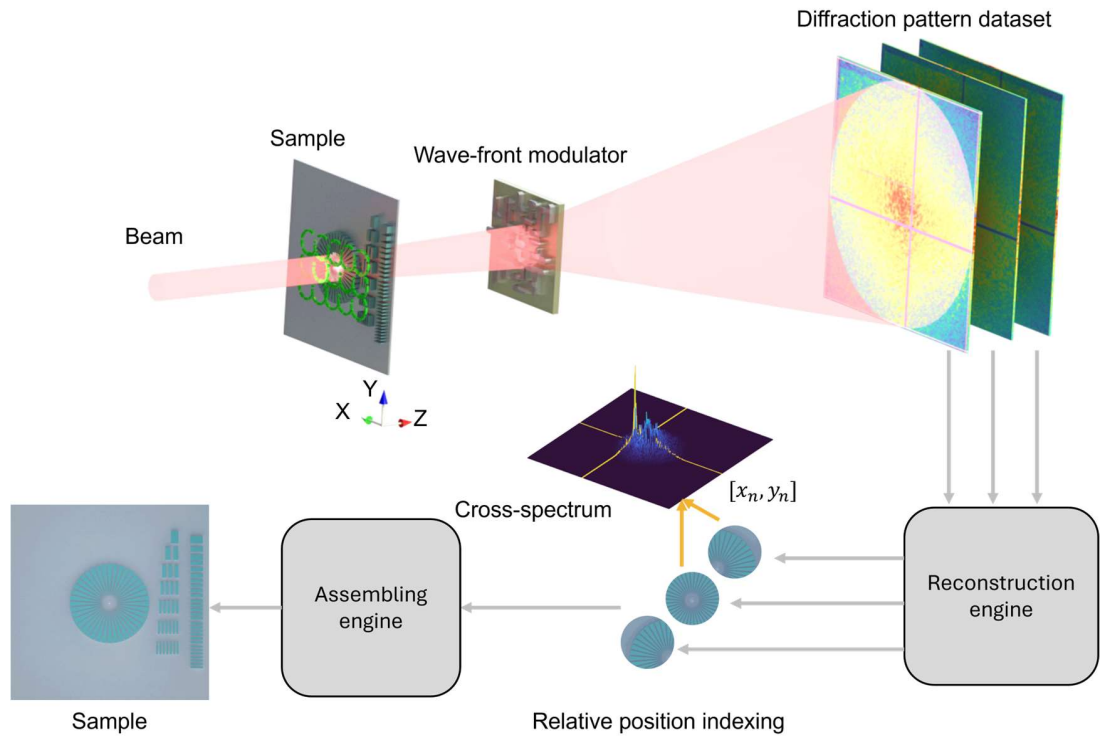

**Fig. 1. Apparatus and workflow of scanning diffraction microscopy without prior knowledge of position.** The probe scans the sample with a certain overlap ratio. Then, exit waves are modulated by the wavefront modulator placed downstream of the sample. The detector records the diffraction intensity dataset. During the reconstruction engine, object functions of all diffraction patterns are recovered in parallel; the relative scan positions are searched by indexing the peak of the cross-spectrum of object functions. After that, the object functions are assembled and refined to form the completed sample function by the assembling engine.

## 2. Results

The method was first validated in an optical experiment on biological samples and then with synchrotron X-ray data. Specifically, the optical demonstration validated the feasibility and robustness against low overlap ratios. The X-ray experiment demonstrates the feasibility of measuring the spatial deviations of probe function under low overlap ratio situations previously inaccessible by other methods.

### A. Optical validation with various overlap ratios

Imaging biological samples facilitates our understanding of cellular structure and extracellular organization. To validate the feasibility of our method, we performed an optical experiment on the wing of a fly. One of the diffraction intensities captured by the detector is shown in Fig.2j. The optical microscope image of the sample is illustrated in Fig.2c. The amplitude and phase of the sample, recovered under 58% of the overlap ratio, are shown in Fig.2a and b. The scaler bar of the image is 350μm. The part surrounded by the black box of Fig.2b was unwrapped using the graph-cutting-based technique [52], yielding the image in Fig.2 f. All surface structures and burrs of the wing are clearly resolved. Visually, these recovered results show similarities with the microscopy image.

Such an agreement can also be observed in low overlap ratio results, shown in Fig.2d, e, g, and h. The low overlap datasets were obtained using a portion of diffraction data from the Fig.2 a and b datasets. According to the previous research on the overlap ratio requirement of Ptychography, 60% of the overlap ratio is typically needed to yield a good image quality [30,41]. However, the modulation mechanism in our method serves as an additional constraint that assists the convergence of iterative reconstruction, reducing the dependence on the overlap ratio. As a result, the proposed method can provide high-quality images in low overlap ratio situations, which are not accessible by traditional Ptychography. Image degradation can hardly be observed even when the overlap ratio is reduced to 13%, as shown in Fig.2g and h.

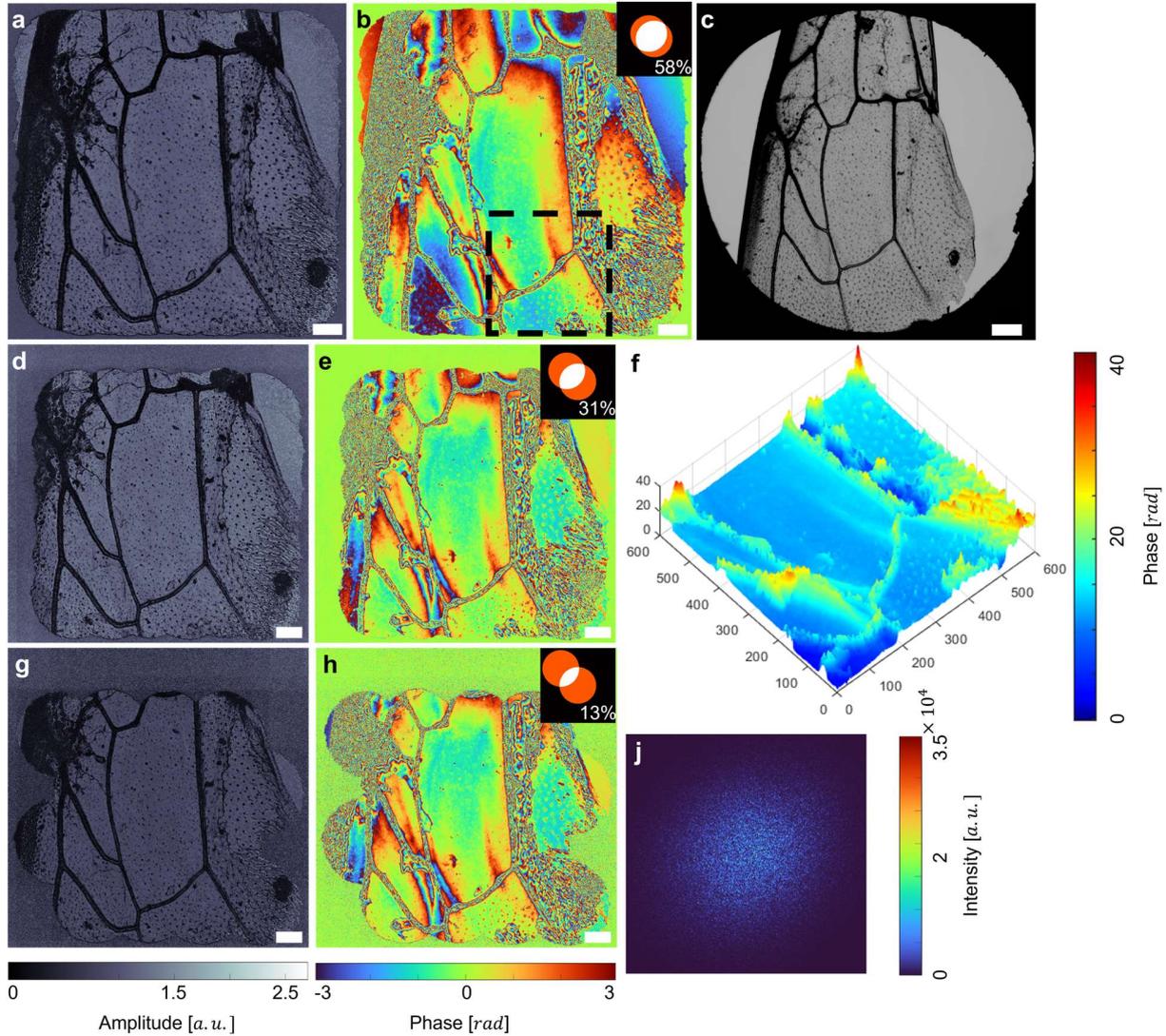

**Fig. 2. Results of optical experiment. a, b, d, e, g, and h** are Reconstructed amplitude and phase of the wing without prior position information under various overlap ratios. **c** Image of optical microscopy. **f** Three-dimensional phase profile of **a**. **j** Exemplary diffraction intensity. The scale bar is 350μm.

Besides the visual fidelity, quantitative analysis of optical results is presented in Fig.3. Fig. 3a shows the plot of the overlap ratio versus position error of the experiment; it is clear that the position accuracy of the proposed method correlates with the overlap ratio, lower overlap ratio results in higher position error which we believe attribute to the lack of reference feature inside the overlap region. The sampling interval under this experimental geometry was 2.9μm. What should be noticed is that the proposed method can still realize sub-pixel accuracy of positioning even when the overlap ratio is reduced to 13%. When the overlap ratio increases to 58%, the difference between the position by the proposed method and state-of-the-art positioning device reduces significantly; the error vector of this overlap ratio is magnified 30000 times and shown in Fig. 3b. The position error along both X and Y axes for most scan points are less than 10nm indicating the position recovery by our method is comparable with the commercial motion stage. Additionally, the distribution of the error vector is not correlated with the raster scan trajectory, indicating that the recovered position by our method does not depend on the scan trajectory but on the sample feature. The reference position of this experiment was obtained from the readout positions of the Newport MFA-CC stage, whose nominal bidirectional repetition accuracy was 150nm. The optical experiment validates the feasibility of imaging and positioning utilizing the shared sample feature of overlapped sample functions. The completed description of experimental parameters is detailed in Method and Materials. Details on the definition and calculation of the overlap ratio can be found in supplementary information. This experiment validated the feasibility and demonstrated its robustness in position recovery for low overlap ratio situations. Such validation provides a possibility for high-throughput and high-resolution diffraction imaging.

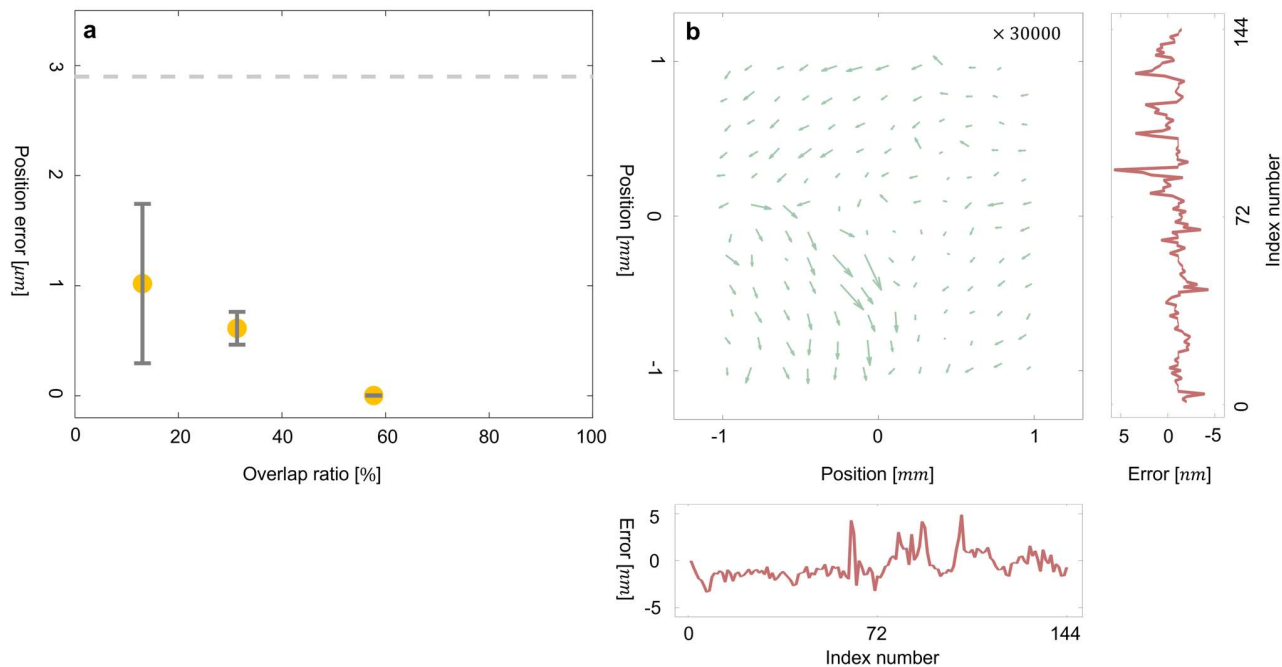

**Fig. 3. Quantitative analysis of the optical experiment. a** Plot of retrieved position error versus overlap ratio. The error bar represents the standard deviation of position error overall scan points. **b** The vectorial distribution of position error vectors magnified by 30000; and error plot along X and Y axes, respectively.

## B. X-ray experiment with spatially unstable probe

High-resolution ptychographic imaging typically employs a nanometer-sized probe, imposing stringent stability requirements on the source and focusing components to meet the stable probe assumption of ptychography. Various techniques have been proposed to compensate for probe variations, including Mixed-state, orthogonal probe relaxation, and optimization methods [49-51]. These techniques have been proven to be useful in tolerating minor probe variations. Pre-calibration and sophisticated control systems are generally needed to guarantee minor probe variations, especially for electronic microscopy. Importantly, calibration procedures vary across systems, and manual calibrations maintain stability only for a limited period, significantly increasing maintenance costs and limiting broader application. In traditional ptychographic engines such as the ePIE and DM, the probe and object functions are separated during each iteration; dramatic deviation of the shared and stable probe assumption will lead to the failure of reconstruction.

In our method, exit waves at each scan position can be recovered without separating the probe and object functions. Therefore, the wavefront recovery process is independent of the scan, providing tolerance for spatial and temporal variations, especially probe variation. We applied the proposed data processing method to our experiment data previously acquired at the I13 beamline, Diamond Light Source, to verify such property. The SEM image of a custom-design test sample is shown in Fig.4 c; The amplitude and phase characterized by our method are shown in Fig. 4a and 4b. The elements and characters are resolved and share similarities with the SEM image. In the reconstruction process, exit waves of all diffraction patterns were first reconstructed without separating the probe and object functions. Exemplary exit waves are shown in Fig. 4 g; the spatial variations of the probe can be clearly observed from their positions inside the support region. The completed exit waves can be found in the Supplementary Information section. The support region is fixed for all the exit waves and marked by the blue dotted circles. Exploiting the shared features of the probe function in recovered exit waves, the deviations of the probe can be quantitatively calculated by the registration algorithm. Beam deviation of 500nm along the X axis and 50nm along the Y axis were measured and shown in Fig.4 f. After compensating for the probe deviation and applying average and subtraction operations on the exit waves, the object functions corresponding to each diffraction pattern are revealed. The registration algorithm determines the relative scan position of the motion stage from the recovered object functions by the registration algorithm due to the overlapping nature of the scan. The recovered scan information of the sample is shown in Fig.4 d, where the orange dots represent the scan points recovered, and the gray circles indicate the size of the probe. The overlap ratio of two points surrounded by the red box is sketched in Fig.4 e, where the green dotted circles represent the region of each wavefront. A detailed description of the experiment is stated in the Method and Materials section.

In this experiment, the deviation of the beam violated the invariable probe assumption of ptychography, and the sample was scanned with only an overlap ratio, by recovering exit waves for each scan point and measuring the relative position numerically while leaving the probe-object separation process independent from reconstruction. We not only recovered the sample and probe functions with high fidelity but also measured the deviation of probe functions regardless of the low overlap ratio. Also, our method effectively handles the ambiguity between probe drift and sample movement, which traditional position refinement algorithms would struggle with. Because the displacement of the illumination region of exit waves indicates the deviation of the probe, after compensating for the deviation of the probe, the shift of the shared feature of overlapped waves is only attributed to the movement of the sample. In addition to the demonstration of quantifying and compensating the drift effect of the beam, it is also possible to quantitatively measure and compensate for other variations in the experiment, such as intensity and phase fluctuations, coherence degradation, and motion blur, by adding additional constraints to the proposed reconstruction workflow. Furthermore, our method's unique capability to recover the instability of every single pulse could provide a quantitative indication of beam quality. This feature is crucial for instrumental design, construction, diagnosis, and systematic maintenance.

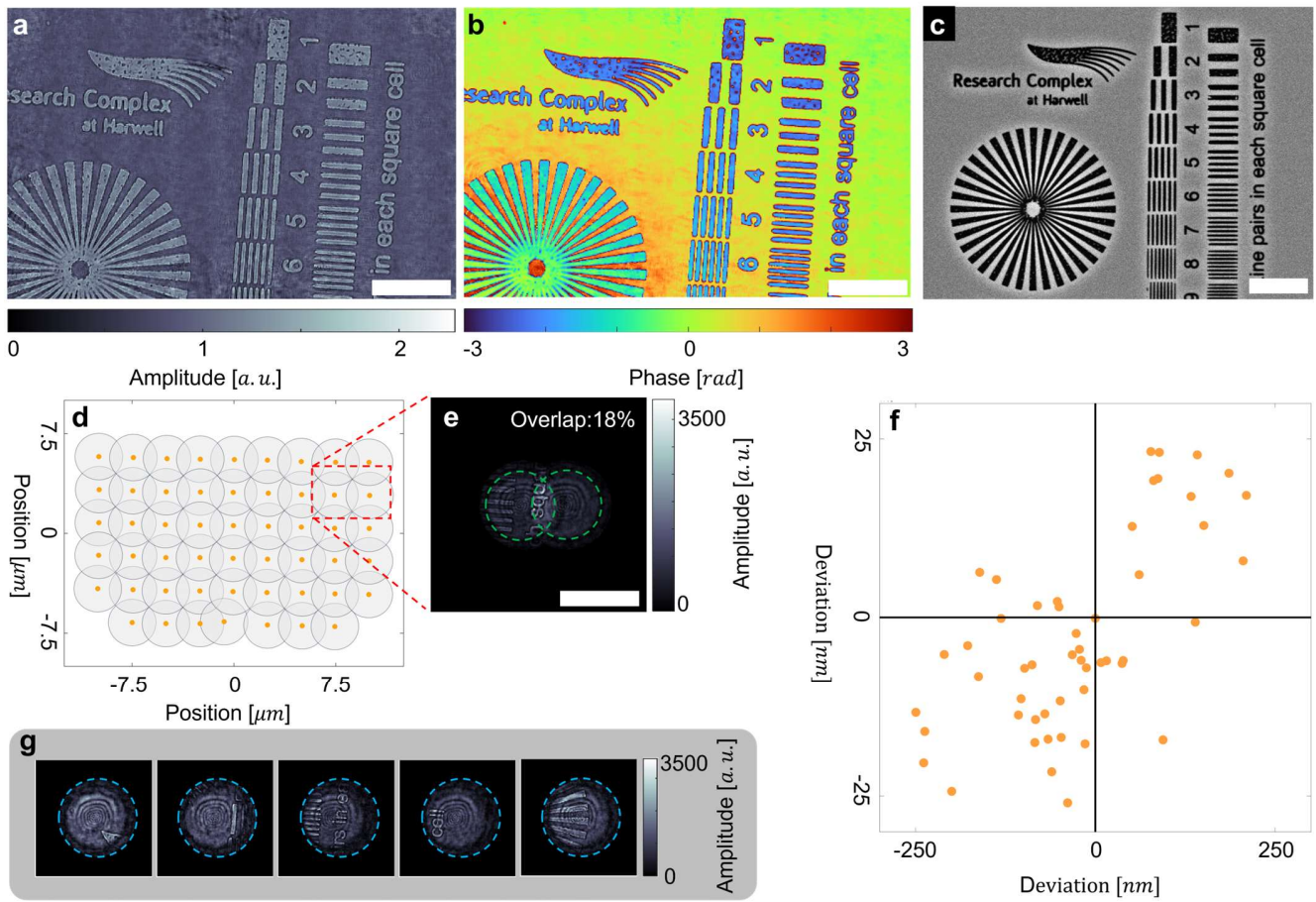

**Fig. 4. Results of X-ray experiment with a low overlap ratio and unstable probe. a, b** Reconstructed amplitude and phase of a customized sample. **c** SEM image of the customized sample. **d** Position map of reconstructed positions, where the orange points and gray circles represent the recovered positions and the probe footprint. **e** Sketch of overlap ratio for the points surrounded by the red box. The green circles indicate the illumination regions. **f** The scatter map of probe drift along the X and Y axes. **g** Exemplary recovered exit-waves with a spatially varying illumination probe. Blue dotted circles represent the region of the applied support constraint. The scale bar is $4\mu m$.

## 3. Discussion

### A. Relationship between retrieved position accuracy and overlap ratio

The proposed method recovers the relative position utilizing the shared feature of object functions. Its position accuracy is thus dependent on the overlap ratio; structure-rich features between two adjacent object functions permit higher positioning accuracy.

We conducted simulations with different sample features and overlap ratios to study the relationship among these factors. The simulation results are shown in Fig.5. In our simulation, we employed two types of samples: one biological sample from our optical experiment and a maze sample; their performance is slightly different due to the characteristics of their features. The maze sample has more distinguishable and fine features inside the overlapped region; therefore, the minimum overlap ratio for successful reconstruction is 3.7%; also, the position error for all scan points is still lower than one pixel. For biological samples, the absorption and sparse features in overlapped regions limit the positioning performance; thus, the minimum achievable overlap ratio is 6.6%, and the error bar of position error is slightly higher than one pixel. When the overlap ratio increases to 10%, both samples can realize subpixel accuracy in position recovery, which is far less than the 60% overlap requirement of Ptychography. The position error reduces significantly for an overlap ratio larger than 25%, demonstrating that our method has deviated from the traditional Ptychography regarding the overlap ratio requirement. Because the recovered position is accurate enough, probe and modulator functions are well known, the samples are recovered with good quality, and they are similar to the original images; therefore, we only show the original sample images in Fig.5.

The periodic structure of a sample could introduce artifacts in traditional Ptychography, which can be alleviated by using structural illumination or a specific scan trajectory [27,30,53-55]. For the proposed method, the periodic structure of the sample can introduce ambiguity in positioning. However, it can be handled if the probe and scan step are smaller than the sample's structure period, and a structural scan trajectory can also help improve the sampling diversity and break the ambiguity introduced by the periodic structure.

Besides, the existing periodic signals in the diffraction pattern introduced by the focusing component might cause misleading artifacts in the image, as the Ptychographic model does not include any mechanism to avoid such artifacts. In contrast, for our method, the non-periodic modulation structure can eliminate the potential periodicity of diffraction signals. These properties have been validated with experimental data, the details of which can be found in the Supplementary Information section.

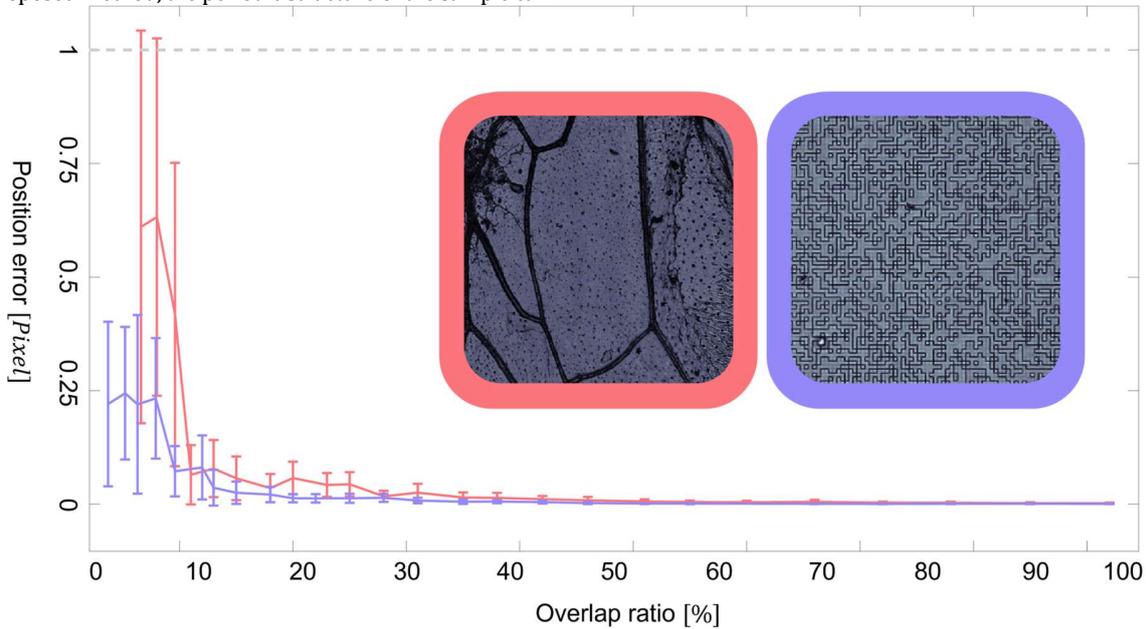

**Fig. 5. Dependence of position accuracy versus different overlap ratios.** The error bar represents the standard deviation of absolute position error overall scan points.

### B. Calibration of the modulator

The proposed method relies on the modulation mechanism introduced by the modulator placed downstream of the sample. Therefore, the modulator transmission function should be calibrated in advance or during the reconstruction. The calibration can be accomplished by conventional Ptychography, as reported in our previous work [47]. Ptychography-based calibration method requires prior knowledge of scanning position, which is difficult to guarantee on an atomic scale. Therefore, we develop a novel modulator calibration procedure different from ptychography. The proposed calibration procedure involves illuminating the modulator with varying structural wavefront and specific update formulas. In the literature, a highly divergent beam could eliminate the ambiguity in the iterative phase retrieval process due to the degradation of the Fraunhofer diffraction pattern [12]. Therefore, a divergent probe was used in our calibration experiment to enhance the convergency. Moreover, the random momentum of photons would increase the sampling diversity, so we employed a diffuser in an experiment to alternate the photon momentum. Varying structural illumination was introduced by lateral translation of the diffuser; other approaches to generating structural beams are also applicable. Besides, the nature of modulator calibration is to recover its transmission function. If the modulation pattern has a periodic structure and its period is large enough to be sufficiently sampled, the existing periodicity of the modulation pattern would allow the modulator function to be sampled more easily.

Additionally, the role of probe and modulator functions are similar in the iterative process because they are shared for all the diffraction patterns. Such ambiguity may lead to the failure of the modulator calibration, but it can be eliminated by providing a coarse prior knowledge of the probe. Considering these principles, we conducted a modulator calibration experiment using a synchrotron X-ray source. Results are shown in Fig.6. A highly focused probe was employed; its cross-section is illustrated in Fig.6 c. The axial position of the diffuser and modulator are indicated by orange and green dot lines, respectively. The known probe function is shown in Fig.6 a. Exemplary reconstructed amplitudes of the diffuser with an area of 1.8μm×1.8μm are shown in Fig.6 b. Importantly, no overlapped features can be observed from these sample exit waves, and their initial estimates are unity matrix. The modulator's SEM image and recovered phase profile are shown in Fig.6 d and e. The averaged period measured by red lines is 388nm and 379nm, respectively, which verifies that the modulator is calibrated accurately, considering the sampling interval of the modulator is 7.1nm and error by manual measurement.

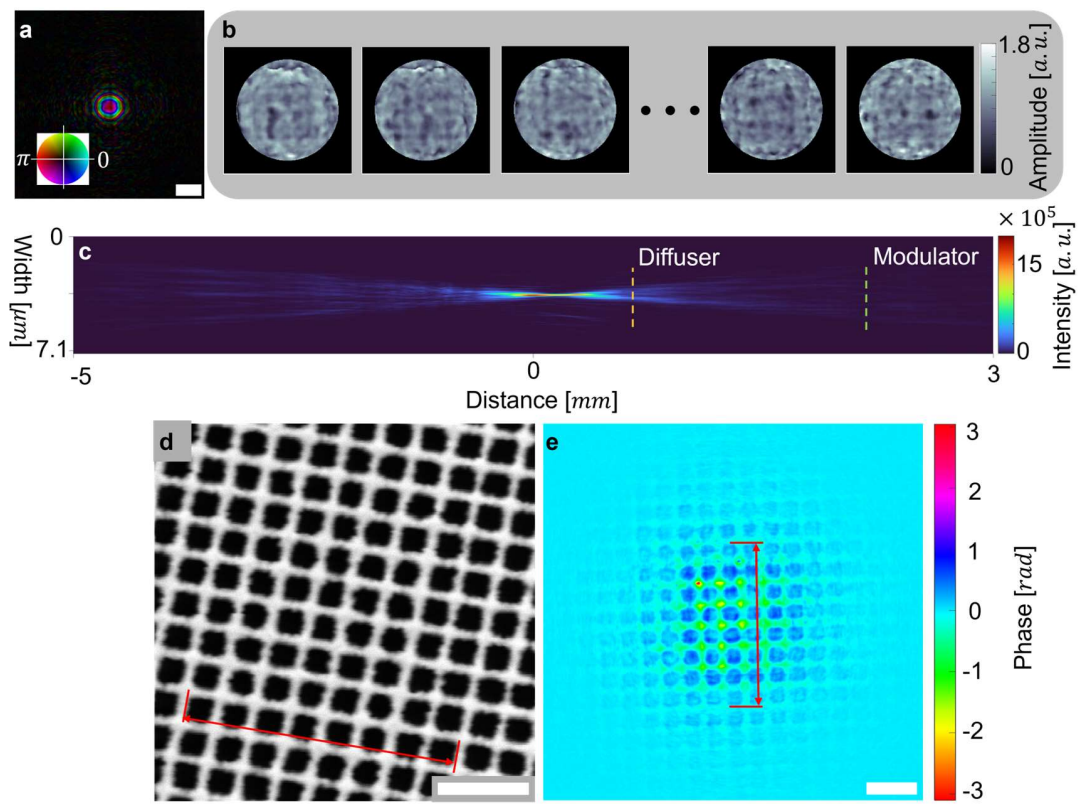

**Fig. 6. Experimental validation of a novel modulator calibration procedure. a** The known probe function is in which the hue represents the phase. **b** Reconstructed amplitude of parts of a diffuser with an initial guess of one and a field of view of 1.8μm×1.8μm. **c** Cross-profile of the probe intensity. Dash lines indicate the position of the scanned diffuser and fixed modulator. **d** SEM image of the grating modulator with an averaged period measured by the red line is 379 nm. **e** Reconstructed phase profile of modulator with an initial guess of one. The averaged period by red line is 388nm, and the pixel size of the reconstructed modulator is 7.1 nm. The scale bar of this figure is 1 μm.

### C. Potential extensibility
Three-dimensional imaging

Understanding thick samples' three-dimensional electron density distribution is crucial for studying their properties. However, modeling dynamic scattering within such samples presents significant challenges. The proposed method recovers only the exit waves of the sample without modeling the interaction between the photon and matter, providing an accessible way to inverse the dynamic scattering process numerically. Specifically, with the complex wavefields obtained, the indexing can be finished in real space, and the dynamic scattering process can be studied through exit waves from different incident orientations inaccessible by previous methods.

Additionally, structural illumination with diverse photon momentum has been proven efficient in enhancing resolution [56,57]. The downstream placement of the wavefront modulator in our approach reserves the possibility of it. In summary, our method holds promise for advancing three-dimensional imaging, particularly in thick samples when dynamic scattering plays a pivotal role.

## 4. Methods and Materials

### A. Details of experiments and simulations
A1. Visible light experiment

In our visible light experiment, the probe was formed by illuminating an aperture with a 632.8nm collimated beam. The aperture with diameters of 0.8mm was positioned 0.15mm upstream of the sample. A commercially available phase-type modulator manufactured by Luminit was employed and placed 11.5mm downstream of the sample. A 30mm focal length lens was inserted between the sample and the Dhyana 400BSI V3 sCMOS detector to emulate the Fraunhofer diffraction recording conditions. The sample was mounted on an assembly of Newport MFA-CC stage with a nominal bidirectional accuracy of 150 nm. The scan trajectory was a raster grid. The completed dataset is composed of 144 diffraction patterns. Moreover, a low overlap ratio dataset was obtained using the original dataset's sparse grid points.

A2. X-ray experiment with spatially unstable probe
The X-ray experiment was conducted at the I13 beamline, Diamond Light Source. The probe was generated by a 7.5KeV collimated beam illuminating an aperture with a diameter of 4.1μm. The aperture was placed 1.7mm upstream of the sample, and a phase-type modulator with a random pattern was placed 6.56mm downstream to introduce sufficient modulation. Diffraction intensity was collected by a Merlin detector located 7.7m from the modulator.

A3. Simulations of position accuracy versus different overlap ratio
The probe in our simulations was formed by illuminating a 1mm aperture with a 632.8nm beam. The aperture was 1mm upstream of the sample. The modulator employed was the same as the optical experiment. The modulator was placed 11.5mm downstream of the sample. The distance between the modulator and the detector was 30mm; propagation between these planes was Fraunhofer diffraction. The simulated detector pixel size was 6.5 μm. A biological sample was recovered in the optical experiment, and a sample with a maze pattern was used as the object in the simulation.

A4. X-ray experiment for modulator calibration
The modulator calibration experiment was implemented with data recorded at the MAXIV NanoMAX beamline. The probe was formed by focusing an 8 KeV collimated beam with a Kirkpatrick–Baez (KB) mirror. The scanned diffuser was placed downstream of the focus. The detector was positioned away from the focus. A phase-type grating pattern modulator was placed downstream of the diffuser and kept fixed during the scanning. The diffuser was moved by Motion Solutions NPXY100Z100-135 sample scanner with a position noise along all the axes.

**B. Reconstruction workflow**
The reconstruction workflow consists of three steps: reconstruction of wavefronts, determination of relative position and sample assembly.

**Reconstruction of wavefronts.** In this part, the shared probe function and object functions of each diffraction pattern are reconstructed in parallel using the following algorithm.

| Algorithm: reconstruction of wavefronts |
|---|
| Input: Diffraction pattern set $I(q_n)$, initial estimate of probe function $P(r)$, support function $S(r)$, object functions $O(r_n)$ and modulator functions $M(r)$. Here, $r$ denotes the spatial coordinate vector, $n$ denotes the diffraction pattern set $\{1,2,3,…n\}$. |
| Output: Revised estimate of probe function $P(r)$, object functions $O(r_n)$, and modulator functions $M(r)$. |

1     Initiate probe $P^j(r)$ and object functions
2     **for** j from 1 to total iterations **do**
3         **if** probe and object separation **then**
4             calculate the exit waves by dot product of probe and object functions.
5         **else**
6             Apply the support constraint on the exit waves.
7         **end**
8         Propagate the exit waves from object plane to front side of modulator plane.
9         Apply modulation.
10        Propagate the modulated exit waves from the rear side of modulator plane to the detector plane.
11        Revise the waves through modulus constraint.
12        Back propagate the exit waves from the detector plane to the rear side of the modulator plane.
13        Update modulator function with specific formula.
14        Undo modulation.
15        Propagate the revised exit waves from the front side of the modulator plane to the object plane.
16        **if** probe and object separation **then**
17            Probe and object functions are updated with averaged update formula.
18        **else**
19            Update the exit waves.
20        **end**
21
22 **end**

In step 13, the formula for updating the modulator transmission function is a modified ePIE equation. To exploit the diversity of modulated wavefronts and the shared nature of modulator function, we modify the conventional ePIE equation by adding an average operation to the wavefronts. Explicitly, the averaged operation is implemented on the wavefronts to extract the information of modulator from the wavefronts. Such an operation is also applied in step 17 to update the shared probe function. In contrast, the object functions are different from each other, so, the traditional ePIE formula is used to update the object function estimate.

For cases when the assumption of a spatially stable probe is not valid, the conventional update formula can't be used. It is necessary to recover the exit waves of each position rather than separating the probe and object functions. Before positioning the object position, a measurement and compensation of probe drift is needed which is realized by aligning all the exit waves. After compensating the drift of probe, the object functions of different positions can be obtained by using formula in step 17. It is also possible to incorporate of new constraints that separating the unstable probes and sample functions.

**Relative position index.** Due to the overlapping nature of the scan, the relative position of all scan points can be found by registering the object functions. In this paper, for the sake of computational

efficiency, we adopt the cross-correlation for registration with formula stated as follows:

$$CS = \int_{-\infty}^{+\infty} f(r)\, g(r - \Delta r) dr \quad (1)$$

The sign $CS$ here represents the cross-spectrum of function $f(r)$ and $g(r)$. However, in experiment, the conventional cross-correlation generally suffers from the deterioration of noise and requirement of high overlap ratio, a feature enhancing technique may be helpful for position registration. To enhance the image feature and harness the characteristic of in-plane translation of ptychography, we propose a scaling gradient enhanced cross-correlation method that includes a gradient precondition technique based on the scaling property of Fourier-transform which is stated as follows:

$$f(ar) \leftrightarrow \frac{1}{|a|} \mathcal{F}\left(j\frac{q}{a}\right)$$
$$f'(r) = f(ar) - f(r) \quad (2)$$

Here, $a$ is the scaling factor; $r$ and $q$ are the real and reciprocal space coordinates, respectively. By subtracting the original object function, the scaling gradient can be obtained. For the discrete calculation, the scaling is realized by interpolation in reciprocal space. The size of the interpolated matrix should be smaller than the original matrix ensuring that the image is enlarged in real space. The scaled image is then padded zero to ensure its size is identical to the original image so as to perform the subtraction operation. The relative positions between the overlapped object functions can be found by applying this processing to the recovered object functions. In addition to the preconditioning method presented here, other strategies, such as extracting the amplitude or phase, might be helpful for the search for relative position. For the sake of computational efficiency, we only demonstrate the feasibility of using cross-correlation to measure the relative position; other image registration techniques can also be applied.

**Assembling engine.** With the information obtained in the above section, many algorithms can be used to assemble the completed sample information. Even traditional ptychography engines, such as the extended ptychography iterative engine (ePIE), difference map (DM), or maximum-likelihood algorithm, can also be used [30,48]. For simplicity, the assembly engine used in our optical and X-ray experiments is ePIE.

### C. Manufacture of modulator

**Material of modulator.** The modulator is essential for the proposed method; therefore, specific consideration of the modulator is needed for different scenarios. According to our previous work, the phase-type modulator with phase shift is appreciated due to its strong performance in alternating wavefront and low energy absorption. For optical wavelength, the modulator used in the experiment is a commercial random pattern phase diffuser by Luminit. The phase-type modulator is manufactured for our X-ray experiments by etching the tungsten for a given depth to realize the designed phase delay. The tungsten substrate is a silicon nitride membrane with a thickness of 200 nanometers. The material of our modulator is designed and optimized for 6-8 KeV; for other energy levels, further consideration of material is needed.

**Modulation pattern design.** The spatial distribution of phase delay contributes to the convergence of the modulator. Spatial distribution can be quantitatively defined by feature size, spatial density, and period. Specifically, feature size denotes the size of the minimum phase retardance unit. Spatial density represents the density of the feature unit per unit area. Both factors jointly contribute to the frequency spectrum of the modulator; the smaller the feature size, the higher the diffraction angle, and the better the modulation effect. The higher the spatial density of the phase delay unit, the higher the average spectrum frequency. Whereas there is a tradeoff between the modulation effect and the difficulty of calibration, a strong scattering modulator would be difficult to calibrate due to its complexity and sensitivity to the system's stability. The difficulty of calibrating the modulator is related to the final parameter - period, which is a key factor controlling the spectrum distribution of the modulator. More specifically, for a given spatial density and feature size, the periodic diffraction intensity introduced by the periodic modulation pattern can be better sampled by a pixelated detector than the random one if its period is large enough to obey the Shannon sampling theorem [58]. The sampling conditions are related to experimental geometry. Therefore, one should consider these three factors when designing a modulator for a specific apparatus.

**Funding.**
This research is supported by the National Natural Science Foundation of China (12074167) and the Shenzhen Science and Technology Innovation Program (KQTD20170810110-313773).

**Acknowledgments.**
The x-ray experiments were performed at the I13-1 beamline, Diamond Light Source, UK, and the NanoMAX beamline, MAXIV, Sweden. We thank I. Robinson, B. Chen, and A. Parsons for their support in the data acquisition.

**Author contributions.**
B. Wang, J. Zhao, and F. Zhang conceived the project. B. Wang conducted the initial simulations. J. Zhao contributed to the early data acquisition. T. Liu designed and conducted experimental validation and data analysis. T. Liu and F. Zhang wrote the manuscript with input from all other authors.

**Disclosures.**
Some of the methods described in this paper are subject to pending patents from the Southern University of Science and Technology.

**Code availability.**
The code of this paper is available at https://github.com/TaoLiu1234/Scanning-diffraction-imaging-without-stable-illumination-and-scan-position-information.git.

**Data availability.**
Data underlying the results presented in this paper are not publicly available at this time but may be obtained from the authors upon reasonable request.

**Supplemental document.**